\documentclass{article}
\usepackage{spconf,amsmath,graphicx}
\usepackage{booktabs}
\usepackage{xcolor}
\usepackage{graphicx}
\usepackage[linesnumbered,ruled,vlined]{algorithm2e}
\SetKwInput{KwInput}{Input}                
\SetKwInput{KwOutput}{Output}              

\title{Enhancing COVID-19 Severity Analysis through Ensemble Methods}
\name{Anand Thyagachandran, Hema A Murthy }
\address{Department of Computer Science and Engineering, \\ Indian Institute of Technology Madras, India \\
\{tanand, hema\}@cse.iitm.ac.in}
\begin{document}
\maketitle

\begin{abstract}
Computed Tomography (CT) scans provide a detailed image of the lungs, allowing clinicians to observe the extent of damage caused by COVID-19. The CT severity score (CTSS) based scoring method is used to identify the extent of lung involvement observed on a CT scan. This paper presents a domain knowledge-based pipeline for extracting regions of infection in COVID-19 patients using a combination of image-processing algorithms and a pre-trained UNET model. Then, an infection rate-based feature vector is proposed for each CT scan. The severity of the infection is then classified into different categories using an ensemble of three machine-learning models: Extreme Gradient Boosting, Extremely Randomized Trees, and Support Vector Machine. The proposed system was evaluated on a validation dataset in the AI-Enabled Medical Image Analysis Workshop and COVID-19 Diagnosis Competition (AI-MIA-COV19D) and achieved a macro F1 score of 64\%. These results demonstrate the potential of combining domain knowledge with machine learning techniques for accurate COVID-19 diagnosis using CT scans. The implementation of the proposed system for severity analysis is available at \textit{https://github.com/aanandt/Enhancing-COVID-19-Severity-Analysis-through-Ensemble-Methods.git }

\end{abstract}
\section{Introduction}
\label{sec:intro}
The outbreak of COVID-19 has resulted in a substantial need for prompt and precise diagnostic testing to detect individuals who may have contracted the SARS-CoV-2 virus. Laboratory tests such as Reverse Transcription Polymerase Chain Reaction (RT-PCR) \cite{emery2004real} and antigen tests are commonly used for diagnosing COVID-19. While these tests detect the virus's presence from the respiratory samples \cite{drame2020should}, but fail to provide an accurate analysis of the disease severity. Additional diagnostic tools are needed to assess the extent of lung damage. Radiological imaging of the chest, including chest radiography and CT scans, is essential to determine the severity of COVID-19 infections \cite{benmalek2021comparing}. At the same time, chest X-rays do not furnish sufficient resolution to evaluate the extent of lung damage. CT scans provide a more detailed view of the lungs and can identify the distribution and area of infection. Clinicians can comprehend critical information to gauge the severity of the disease and deliver timely and effective treatment to the patients.

\begin{table}[h!]

\centering
\caption{Summary of the dataset (COVID-19-CT-DB)}
\begin{tabular}{c |c |c} 

\toprule
Category~ & Train & Validation  \\
\toprule
Mild      & 133   &     31        \\
Moderate  & 124      &    20         \\
Severe    & 166      &     45        \\
Critical  & 39      &     5 \\
\bottomrule
\end{tabular}
\label{tab:dataset}

\end{table}
Radiologists typically observe ground-glass opacities (GGO)  \cite{kwee2020chest}, which indicate lung inflammation but that do not obstruct the underlying pulmonary vessels. Consolidations are the advanced stage of GGO and hide the underlying vessels  \cite{yu2020thin}. Pleural effusion occurs when fluid accumulates excessively in the pleural space surrounding the lungs and is a highly severe case of COVID-19 \cite{nambu2014imaging}. These clinical features are critical in identifying and diagnosing COVID-19 in patients \cite{zhang2020clinical}. In addition to GGO and consolidations, radiologists observe features such as the halo sign (central consolidations surrounded by GGO), the reverse halo sign (central ground-glass lucent area with peripheral consolidation), and crazy paving patterns. These features and their distribution provide crucial information to diagnose and manage COVID-19 patients \cite{kwee2020chest, lin2021ct}.

Machine learning and deep learning methods have been extensively encountered to classify and segment the infection regions from the CT scans. The classification task between COVID-19 and non-COVID-19 is proposed in various studies such as \cite{Hou_2021_ICCV, Kollias_2021_ICCV, Liang_2021_ICCV, Miron_2021_ICCV}. Additionally, some studies have extended the classification task to include three categories - COVID-19, Community-acquired Pneumonia (CAP), and Normal, such as \cite{afshar2021covid, garg2021multi,chaudhary2021detecting, xue2021covid}. COVID-19 and CAP diseases share similar features; distinguishing between these two categories is crucial to monitor the progression of COVID-19, which tends to be much faster than that of CAP. These methods are performed well enough in classification but need to identify the severity of patients. Further, many research works have been focused on the infection region segmentation from CT scans that can be used for severity analysis. Generating large medical image datasets for infection segmentation is time-consuming and requires highly qualified domain experts to annotate the data. Different strategies are proposed to address the unavailability of the large corpus, such as semi-supervised method (small dataset is used for supervised training along with a large amount of unlabelled data) \cite{Fan, ma2020active}, weakly supervised way (ground truth data used partially in training process) \cite{Wang, Han}, and unsupervised methods (without any ground truth data) \cite{9412228, chen2022unsupervised}. A criterion on the volume of infection has been used as an indicator of the severity of infection \cite{ye2021severity, qiblawey2021detection} and classified the CT scans into different classes such as healthy, mild, moderate, severe, and critical.

Radiologists commonly use the CT severity score (CTSS) to determine the extent of severity in COVID-19 patients \cite{lieveld2021chest}. The score ranges from 0 to 25 and is calculated based on the distribution and magnitude of abnormalities (lung involvement) seen on chest CT scans. A higher CTSS score indicates more severe lung damage, and this information is essential in making appropriate treatment decisions \cite{li2020ct, li2020clinical}. In line with this approach, we have developed an automatic method that categorizes CT severity into four classes: mild, moderate, severe, and critical. The proposed method employs various image processing algorithms and a pre-trained UNET model \cite{hofmanninger2020automatic} to extract infected regions from CT scans. Then, an infection rate-based feature vector is proposed for each CT scan. Various classical machine-learning models and ensemble-based models learn these representative feature vectors to predict the severity classes for chest CT scans. A fully automatic severity analysis model can significantly reduce clinicians' time to assess a patient's condition.


The paper is organized as follows: \textbf{Section \ref{sec:dataset}} explains the dataset used in the study. \textbf{Section \ref{sec:proposed-method}} presents the proposed method for segmenting relevant clinical features and classifying COVID-19 patients into different severity classes. \textbf{Section \ref{sec:results}} discusses results and inferences. Finally, \textbf{Section \ref{sec:conclusion}} summarizes the present study.

\begin{table}[]

\caption{Different classes of severity and its characteristics in the given dataset \cite{kollias2023ai}.\\}
\resizebox{0.49\textwidth}{!}{
\begin{tabular}{c|c|c}
\toprule
\textbf{Category} & \textbf{Severity} & \textbf{Description}   \\
\toprule
1               & Mild              & \begin{tabular}[c]{@{}c@{}}Few od no GGOs. Pulmonary parenchymal \\ involvement $ \leq $ 25\% or absence\end{tabular} \\
\hline
2                 & Moderate          & \begin{tabular}[c]{@{}c@{}}Ground glass opacities. Pulmonary parenchymal\\ involvement 25 $ \leq $ 50\%\end{tabular}                   \\
\hline
3                 & Severe            & \begin{tabular}[c]{@{}c@{}}Ground glass opacities. Pulmonary parenchymal\\ involvement 50 $ \leq $ 75\%\end{tabular}                   \\ \hline
4                 & Critical          & \begin{tabular}[c]{@{}c@{}}Ground glass opacities. Pulmonary parenchymal\\ involvement $\geq$ 75\%\end{tabular}    \\
\bottomrule
\end{tabular}
}
\label{tab:dataset_severity}
\end{table}
\section{Dataset}
\label{sec:dataset}

The COVID-19-CT-Database (COVID-19-CT-DB) is provided as part of the "AI-enabled Medical Image Analysis Workshop, and COVID-19 Diagnosis Competition (AI-MIA-COV19D)" \cite{kollias2022ai, arsenos2022large, kollias2021mia, kollias2020deep, kollias2020transparent, kollias2018deep}. The CT scans were collected from September 1, 2020, to March 31, 2021. Four highly experienced medical experts annotated these CT scans (two radiologists and two pulmonologists). 

Generally, CT scans are volumetric data (3D scans) and are usually available in Digital Imaging and Communications in Medicine (DICOM) and Neuroimaging Informatics Technology Initiative (NIfTI) formats. These two image formats represent the CT scan image in higher resolution, also known as  Hounsfield Units (HU). Radiologists use a mapping table based on the HU values to identify the different parts of the CT scan images and lung abnormalities. The COVID-19-CT-DB dataset transforms the HU scale image into an 8-bit gray-scale (JPG) image. This down-sampling process introduces an information loss and reduction in the resolution of the CT scan images. 

The severity assessment challenge consists of 462 CT scans for training (456 patients' CT scans are used for training models) and 101 CT scans for validating the robustness of the trained model. The number of slices in each patient varies from a minimum of 27 to a maximum of 702 slices. The summary of the severity analysis dataset is shown in \textbf{Table \ref{tab:dataset}}, and the description of severity categories are given in \textbf{Table \ref{tab:dataset_severity}}. The test data set consists of 231 CT scans. Five patients' data in the test dataset consists of coronal slices of the CT scan. The patient's ids for those CT scans are 'test\_ct\_scan\_138', 'test\_ct\_scan\_139', 'test\_ct\_scan\_140', 'test\_ct\_scan\_141', and 'test\_ct\_scan\_197'. The training and validation dataset did not contain coronal slices of CT scans.

\section{Proposed Method}
\label{sec:proposed-method}

The proposed pipeline consists of two parts, lesion segmentation from the CT scan and quantifying the severity based on the infection rate. The proposed method uses various image processing algorithms, machine learning, and deep learning models to predict the severity score for COVID-19 patients.


\subsection{Infection Segmentation}
The infection segmentation pipeline uses domain knowledge to extract the lesion areas. Initially, the lung region is extracted from the chest CT image by a pre-trained UNET model. Then, the resultant image's contrast is enhanced by histogram hyperbolization \cite{frei1977image}. Next,  the pulmonary blood vessels are enhanced using the top-hat morphological operation \cite{zhu2012automatic}. Finally, the resultant image combines morphological operations to fine-tune the infection regions. The complete overview of the pipeline is shown in \textbf{ Algorithm \ref{alg:algo}}.
\begin{algorithm}[h!]
\label{alg:algo}
\caption{The proposed algorithm to extract the infection regions from CT scans}
\DontPrintSemicolon
   \footnotesize
    \KwInput{CT Scan (3D Data) -- V}
    \KwOutput{Stacked infection regions from CT scans -- S}
    \For{ Image in $V$} 
    {
        $Lung\_mask$ $\gets$ $UNET(HU\_Image(Image))$
        
        $Seg\_img$ $\gets$ {$Lung\_mask$ * $Image$}
        
        \If{(Sum($Lung\_mask$) $\geq$ 10000 and Position(Image) $\geq$ Len(V)/3 and Position(Image) $\leq$ Len(V) * 2/3) or (Sum($Lung\_mask$) $\geq$ (0.7 * (512 * 512)))}
        {
            $Mask_1$ $\gets$ {$Threshold\_intensity\_filter(Seg\_img)$}\\
           $Seg\_img$ $\gets$ {$Mask_1$ * $Seg\_img$}\\
             $CT\_hyper$ $\gets$ {$Hyperbolization (Guassian\_filter (Seg\_img))$}\\
             $CT\_hyper\_new$ $\gets$ {$CC\_area\_filter(MO\_open (CT\_hyper))$}\\
            $Vessel\_mask$ $\gets$ {$CC\_area\_filter(Bin\_OTSU(Top\_hat(CT\_hyper)))$}\\
           $temp\_mask$ $\gets$ {$CT\_hyper\_new$ - $Vessel\_mask$}\\
             $Infection\_mask$ $\gets$ {$MO\_dilate(Fill\_holes(temp\_mask))$}\\
           $S$ $\gets$ $Infection\_mask$\\
        }
        \Else
        {
    	{ Remove the 2D image from further analysis}
        }
   } 

  \tcc{CC:- Connected Component, MO:- Morphological operation}
  
\end{algorithm}

\subsubsection{Lung Mask Generation}
The chest CT scan contains the lung region and organs like the trachea, diaphragm, heart, stomach, and tissues. This module aims to extract the lung region from the chest CT scan and remove the unnecessary areas from further analysis. A pre-trained UNET \cite{hofmanninger2020automatic} model is used to extract the lung region from the CT scan images. This model is trained on the HU scale CT images, and the images in the given dataset are in JPG format. A linear transformation is applied to the JPG images to make them compatible with the input format for the UNET model.


The UNET model is an end-to-end fully convolution neural network comprising a contraction and expansion path. The contracting path consists of 2D convolution layers, a nonlinear activation function, and average pooling layers. A high-dimensional feature map is finally generated. This feature map is up-sampled (by using transposed convolution) and concatenated with the feature map generated from the contracting path through skip connections. These concatenated feature maps are passed through 2D convolution layers and nonlinear activation functions in the expanding path to produce a segmentation mask from the feature map. This is primarily an encoder-decoder model. 
\begin{figure}[t]
    \centering
    \includegraphics[width=0.48\textwidth]{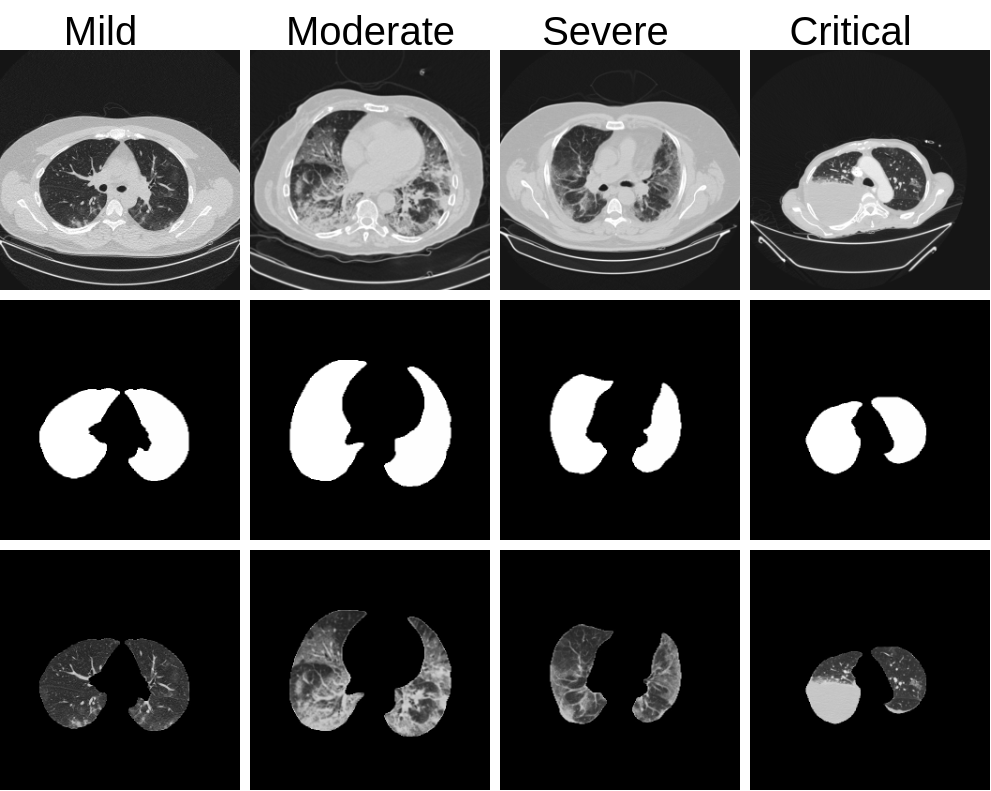}
 
    \caption{First row represents CT scan images of mild, moderate, severe, and critical categories provided in the challenge. The second row shows the lung mask generated from the images. The third row depicts the segmented region of interest for further analysis of each category.}
    \label{fig:lungmask}

\end{figure}

The lung mask involvement accounts for variations in the number of CT image slices across patients. An empirical threshold on the lung mask's involvement and a threshold on the number of slices from the middle region (slice number which lies between one-third to two-thirds of the total number of slices) are used to select the CT scan images for further analysis. The region of interest is extracted using the lung mask from the chest CT scan images. The segmented CT scan images are applied with a  Gaussian filter to smooth the image without significantly reducing the sharpness of the edges. Further, this image is fed to the image enhancement module. The lung masks and extracted regions of interest from different categories of COVID-19 severity are shown in \textbf{Fig. \ref{fig:lungmask}}.

\begin{figure}[t]
    \centering
    \includegraphics[width=0.48\textwidth]{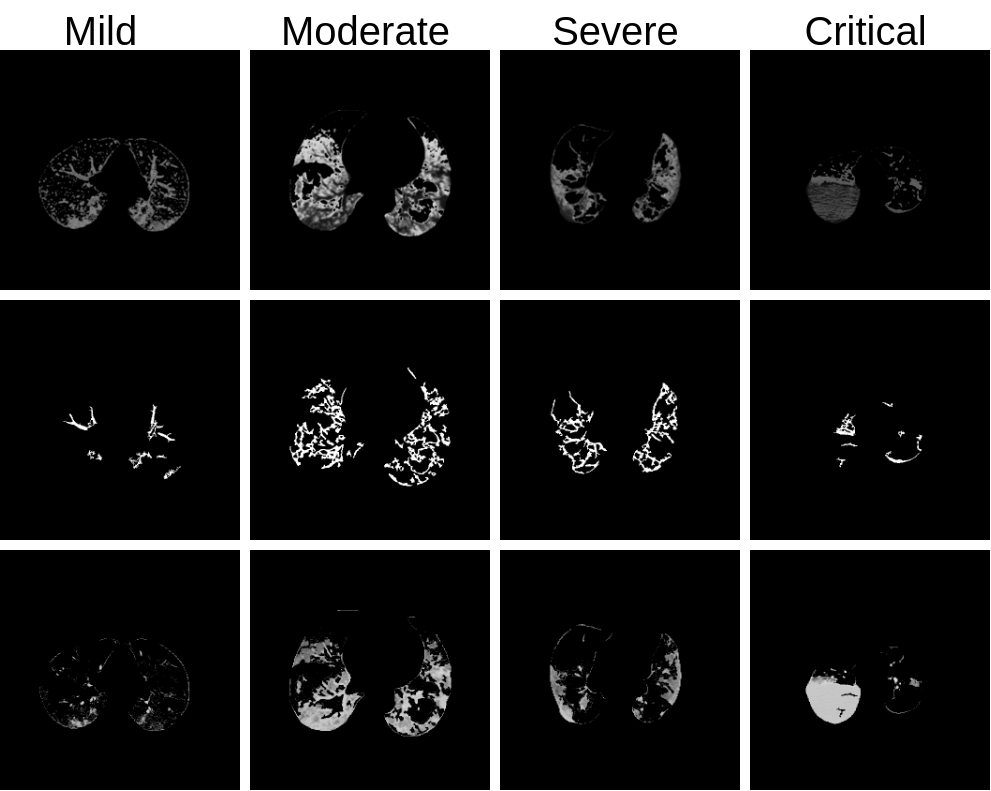}
    
    \caption{First row shows the histogram hyperbolized lung region of mild, moderate, severe, and critical classes. The second row depicts enhanced blood vessels using top-hat transform. The third row represents the extracted infection regions.}
    \label{fig:morphologicaloperations}
    
\end{figure} 

\subsubsection{Image Enhancement using Histogram Hyberbolization}
Histogram hyperbolization, introduced in \cite{frei1977image}, is a non-linear image enhancement method that improves the contrast of an image by adjusting its perceived brightness levels. The primary objective of this approach is that it mimics the process of vision and could correspond to the radiologist's view of the image.


    \begin{equation}
        \resizebox{.7\hsize}{!}{$J(I) = c (exp[ log (1 + \frac{1}{c}) * normcm[I]] - 1)$}
    \end{equation}

where $J(I)$ is the space-invariant hyperbolization transformation of the image $I$. $c$ is an empirical threshold. $normcm$ is the normalized cumulative distribution of the intensity histogram. The contrast-enhanced image contains infection regions and the pulmonary vessels. The contrast-enhanced lung region of the CT scan images using histogram hyperbolization is shown in \textbf{Fig. \ref{fig:morphologicaloperations}}.
%

            
            

\subsubsection{Morphological operations}
The present work mainly uses three commonly-used morphological operations: dilation, opening, and top-hat \cite{maragos1987morphological}. A kernel matrix (a rectangular kernel (3 x 3)) is used to convolve with the input image. The dilation operation chooses the maximum pixel value from each (3 x 3) neighborhood and replaces the center pixel. The erosion operation finds the minimum pixel value from the kernel neighborhood. The open morphological procedure removes any small foreground objects from the binary image by employing the erosion operation followed by the dilation operation. The top-hat transformation enhances blood vessel-like structures by subtracting the image from its open morphological image. Foreground-connected components are computed from the resultant image. A criterion on the size or area of involvement is used to retain the connected components.

The OTSU adaptive binarization method  \cite{otsu1979threshold} is applied to the contrast-enhanced image from the previous module. Morphological operation top-hat transform is applied to enhance and remove the blood vessels. The resultant image is then fed to morphological operations open and area-based foreground filter to remove the small connected components in the resultant image. The dilation operation is employed to fine-tune the boundary of the extracted infection region. The enhanced blood vessels and infection masks generated from the different severity classes are shown in  \textbf{Fig. \ref{fig:morphologicaloperations}}.

\subsection{Lesion To Severity Score}
This section describes the various methods explored to identify the correlation between the infection regions in the CT images and their severity class. Inspired by the CTSS method, a weighted average on the percentage of infection in the left and right lungs is estimated. Further, various machine learning algorithms are experimented with to classify the CT scan into different severity classes. 

\subsubsection{Weighted Average Method (WAM)}
The infection rate in the left and right lungs are estimated separately for each slice in the CT scan with the help of the infection mask and the UNET lung mask. In CTSS estimation, the right lung is divided into three lobes and the left into two lobes, and the radiologist visually determines each lobe's degree of involvement. The lobe scores are calculated by the percentage of infection involvement in each lobe \cite{yang2020chest}. An average score lobe across the CT scan slices and sum up to the final CTSS. In the WAM, a score ranging from one to four is assigned based on the rate of infection,  one (0- 25\% ), two (25 - 50\% ), three (50 - 75\% ), and four (75 - 100 \% ). We employed a similar approach to the CTSS, a proportional weighted average method, to find the severity score for each CT scan slice. The right lung region is multiplied by a weight of three and the left lung region by two. The severity score is estimated by averaging the scores across the CT scan images.

\subsubsection{Non-linear Methods}
This study utilizes a feature vector with a dimension of 80 to represent a CT scan. Two features are extracted from each image, such as the left and right lung infection rates. Two methods are employed for a fixed-length representation of the CT scans. If the CT scan contains more than 40 slices, the slices are uniformly divided into forty regions, choosing the median slice features from each region. On the other hand, the feature vector is enhanced by computing the average percentage of infection from the available slices and appending it to generate an 80-dimensional vector. Machine learning models, including logistic regression (LR), gradient boost (Gboost), Ada boost, k-nearest neighbor (k-NN), naive Bayes (NB), random forest (RF), extreme gradient boosting (XGboost) \cite{chen2015xgboost},  extremely randomized tree (ERT) \cite{geurts2006extremely}, support vector machine (SVM), and voting based ensemble models, are employed to learn from these feature representations to predict the severity classes. The machine learning models are implemented with the help of the sklearn library in Python.  

\begin{table*}[]
\caption{The results of severity analysis of linear and non-linear classifiers using CT scans.}
\resizebox{0.99\textwidth}{!}{%
\begin{tabular}{c|c|c|c|c|c|c|c|c|c|c|c|c}
\toprule
\textbf{Models}    & \textbf{WAM} & \textbf{LR} & \textbf{RF} & \textbf{k-NN} & \textbf{NB} & \textbf{Adaboost} & \textbf{Gboost} & \textbf{XGboost} & \textbf{SVM} & \textbf{ERT} & \textbf{Ensemble1} & \textbf{Ensemble2} \\ 
\toprule
\textbf{Precision} & 0.43         & 0.38        & 0.42        & 0.53          & 0.50        & 0.54              & 0.69            & 0.62             & 0.70         & 0.73         & 0.73               & 0.74               \\
\textbf{Recall}    & 0.46         & 0.40        & 0.44        & 0.48          & 0.60        & 0.50              & 0.52            & 0.60             & 0.59         & 0.61         & 0.61               & 0.62               \\ 
\textbf{F1 score}  & 0.43         & 0.38        & 0.43        & 0.47          & 0.47        & 0.49              & 0.54            & 0.60             & 0.61         & 0.64         & 0.63               & 0.64               \\ 
\bottomrule
\end{tabular}
}
\label{tab:resultsclassifiers}
\end{table*}
\begin{figure*}[t]
    \includegraphics[width=\textwidth]{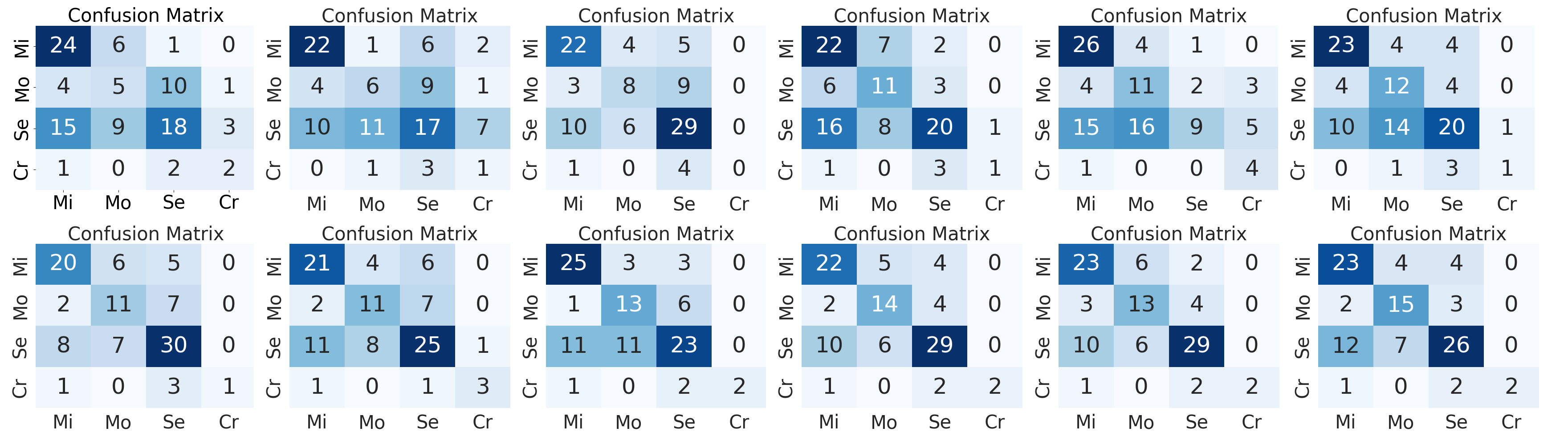}
    \caption{ The first row (from left to right) depicts the confusion matrices of the WAM, LR, RF, k-NN, NB, and Adaboost. The second row represents the Gboost, XGboost, SVM, ERT, Ensemble1 (Adaboost, Gboost, ERT), and Ensemble2 (XGboost, SVM, ERT) classifiers. The axis labels such as Mi-mild, Mo-moderate, Se-severe, and Cr-critical.}
    \label{fig:confusion_matrix}
  
\end{figure*}



\section{Results and Discussions}
\label{sec:results}

The proposed pipeline aims to identify the correlation between the severity class and the lung infection rate. The weighted average method finds a linear mapping between the infection rate and the severity classes, which yielded a macro F1 score of 0.43. This result led to the realization of the non-linear relationship between the infection rate and severity analysis, prompting the exploration of various machine-learning methods. The results of various classifiers are shown in \textbf{Fig. \ref{fig:confusion_matrix}} and tabulated in \textbf{Table \ref{tab:resultsclassifiers}}.

The machine learning models achieve better results (regarding macro F1 score) than the WAM except for logistic regression. The XGboost, ERT, and SVM serve top-3 performing models for the severity analysis. An ensemble of these models achieves the best-performing system in the experiments. The ensemble model can improve the accuracy and robustness of predictions, reduce over-fitting, and improve the model's generalization. The confusion matrices show that the mild class achieves higher classification accuracy and less misclassification with other classes. In the case of the moderate class, the trained models need clarification with the severe classes; the SVM, Ensemble, and RF models provide more misclassifications to the severe category than the mild category. Similarly, the severe class CT scans are misclassified into the mild and moderate classes. While the proposed pipeline works well, it has some class misclassification issues.   While we have observed that the boundary detector has issues in some extreme cases, appropriate domain information is required to improve lung segmentation accuracy. 


The standard evaluation metrics, such as precision, recall, and macro F1 score, are used to evaluate the models. Precision refers to the proportion of true positives among all predicted positive instances. Recall measures the proportion of true positives among all actual positive instances. The Macro F1 score is a harmonic mean of precision and recall, providing an overall measure of model performance across all classes. It considers false positives and negatives and is a valuable metric for imbalanced classes. The performance of linear and non-linear classifiers is shown in \textbf{Table \ref{tab:resultsclassifiers}}.  The baseline model is based on a convolutional neural network (CNN)- recurrent neural network (RNN) where the CNN is used as a feature extractor and RNN is used to model the sequence of features extracted from the CT scan images. The baseline model achieved a macro F1 score of 38\%. It is observed from the \textbf{Table \ref{tab:resultsclassifiers}} that all non-linear models, except logistic regression, outperformed the baseline model. A voting-based ensemble of XGboost, ERT, and SVM  models achieves the best result on the validation data.

\section{Conclusion}
\label{sec:conclusion}

In this paper, we proposed a domain knowledge-based preprocessing pipeline to extract the relevant lesion regions. Infection rate-based features are proposed. Linear and non-linear models are used to predict the CT scan severity class. A voting-based XGboost, ERT, and SVM ensemble model achieves a macro F1 score of 64\%.


\bibliographystyle{IEEEbib}
{
\small
\bibliography{refs}

\begin{thebibliography}{10}

\bibitem{emery2004real}
Shannon~L Emery et~al.,
\newblock ``Real-time reverse transcription--polymerase chain reaction assay
  for sars-associated coronavirus,''
\newblock {\em Emerging infectious diseases}, vol. 10, no. 2, pp. 311, 2004.

\bibitem{drame2020should}
Moustapha Dram{\'e} et~al.,
\newblock ``Should rt-pcr be considered a gold standard in the diagnosis of
  covid-19?,''
\newblock {\em Journal of medical virology}, 2020.

\bibitem{benmalek2021comparing}
Elmehdi Benmalek et~al.,
\newblock ``Comparing ct scan and chest x-ray imaging for covid-19 diagnosis,''
\newblock {\em Biomedical Engineering Advances}, vol. 1, pp. 100003, 2021.

\bibitem{kwee2020chest}
Thomas~C Kwee and Robert~M Kwee,
\newblock ``Chest ct in covid-19: what the radiologist needs to know,''
\newblock {\em Radiographics}, vol. 40, no. 7, pp. 1848--1865, 2020.

\bibitem{yu2020thin}
Minhua Yu et~al.,
\newblock ``Thin-section chest ct imaging of covid-19 pneumonia: a comparison
  between patients with mild and severe disease,''
\newblock {\em Radiology: Cardiothoracic Imaging}, vol. 2, no. 2, pp. e200126,
  2020.

\bibitem{nambu2014imaging}
Atsushi Nambu et~al.,
\newblock ``Imaging of community-acquired pneumonia: roles of imaging
  examinations, imaging diagnosis of specific pathogens and discrimination from
  noninfectious diseases,''
\newblock {\em World Journal of Radiology}, vol. 6, no. 10, pp. 779, 2014.

\bibitem{zhang2020clinical}
Nan Zhang et~al.,
\newblock ``Clinical characteristics and chest ct imaging features of
  critically ill covid-19 patients,''
\newblock {\em European Radiology}, vol. 30, pp. 6151--6160, 2020.

\bibitem{lin2021ct}
Liaoyi Lin et~al.,
\newblock ``Ct manifestations of coronavirus disease (covid-19) pneumonia and
  influenza virus pneumonia: A comparative study,''
\newblock {\em American Journal of Roentgenology}, vol. 216, no. 1, pp. 71--79,
  2021.

\bibitem{Hou_2021_ICCV}
Junlin Hou et~al.,
\newblock ``Cmc-cov19d: Contrastive mixup classification for covid-19
  diagnosis,''
\newblock in {\em Proceedings of the IEEE/CVF International Conference on
  Computer Vision (ICCV) Workshops}, October 2021, pp. 454--461.

\bibitem{Kollias_2021_ICCV}
Dimitrios Kollias et~al.,
\newblock ``Mia-cov19d: Covid-19 detection through 3-d chest ct image
  analysis,''
\newblock in {\em Proceedings of the IEEE/CVF International Conference on
  Computer Vision (ICCV) Workshops}, October 2021, pp. 537--544.

\bibitem{Liang_2021_ICCV}
Shuang Liang et~al.,
\newblock ``A hybrid and fast deep learning framework for covid-19 detection
  via 3d chest ct images,''
\newblock in {\em Proceedings of the IEEE/CVF International Conference on
  Computer Vision (ICCV) Workshops}, October 2021, pp. 508--512.

\bibitem{Miron_2021_ICCV}
Radu Miron et~al.,
\newblock ``Evaluating volumetric and slice-based approaches for covid-19
  detection in chest cts,''
\newblock in {\em Proceedings of the IEEE/CVF International Conference on
  Computer Vision (ICCV) Workshops}, October 2021, pp. 529--536.

\bibitem{afshar2021covid}
Parnian Afshar et~al.,
\newblock ``Covid-ct-md, covid-19 computed tomography scan dataset applicable
  in machine learning and deep learning,''
\newblock {\em Scientific Data}, vol. 8, no. 1, pp. 1--8, 2021.

\bibitem{garg2021multi}
Pratyush Garg et~al.,
\newblock ``Multi-scale residual network for covid-19 diagnosis using
  ct-scans,''
\newblock in {\em ICASSP 2021-2021 IEEE International Conference on Acoustics,
  Speech and Signal Processing (ICASSP)}. IEEE, 2021, pp. 8558--8562.

\bibitem{chaudhary2021detecting}
Shubham Chaudhary et~al.,
\newblock ``Detecting covid-19 and community acquired pneumonia using chest ct
  scan images with deep learning,''
\newblock in {\em ICASSP 2021-2021 IEEE International Conference on Acoustics,
  Speech and Signal Processing (ICASSP)}. IEEE, 2021, pp. 8583--8587.

\bibitem{xue2021covid}
Shuohan Xue and Charith Abhayaratne,
\newblock ``Covid-19 diagnostic using 3d deep transfer learning for
  classification of volumetric computerised tomography chest scans,''
\newblock in {\em ICASSP 2021-2021 IEEE International Conference on Acoustics,
  Speech and Signal Processing (ICASSP)}. IEEE, 2021, pp. 8573--8577.

\bibitem{Fan}
Deng-Ping Fan et~al.,
\newblock ``Inf-net: Automatic covid-19 lung infection segmentation from ct
  images,''
\newblock {\em IEEE Transactions on Medical Imaging}, vol. 39, no. 8, pp.
  2626--2637, 2020.

\bibitem{ma2020active}
Jun Ma et~al.,
\newblock ``Active contour regularized semi-supervised learning for covid-19 ct
  infection segmentation with limited annotations,''
\newblock {\em Physics in Medicine \& Biology}, vol. 65, no. 22, pp. 225034,
  2020.

\bibitem{Wang}
Xinggang Wang et~al.,
\newblock ``A weakly-supervised framework for covid-19 classification and
  lesion localization from chest ct,''
\newblock {\em IEEE Transactions on Medical Imaging}, vol. 39, no. 8, pp.
  2615--2625, 2020.

\bibitem{Han}
Zhongyi Han et~al.,
\newblock ``Accurate screening of covid-19 using attention-based deep 3d
  multiple instance learning,''
\newblock {\em IEEE Transactions on Medical Imaging}, vol. 39, no. 8, pp.
  2584--2594, 2020.

\bibitem{9412228}
Rui Xu et~al.,
\newblock ``Unsupervised detection of pulmonary opacities for computer-aided
  diagnosis of covid-19 on ct images,''
\newblock in {\em 2020 25th International Conference on Pattern Recognition
  (ICPR)}, 2021, pp. 9007--9014.

\bibitem{chen2022unsupervised}
Han Chen et~al.,
\newblock ``Unsupervised domain adaptation based covid-19 ct infection
  segmentation network,''
\newblock {\em Applied Intelligence}, pp. 1--14, 2022.

\bibitem{ye2021severity}
Ben Ye et~al.,
\newblock ``Severity assessment of covid-19 based on feature extraction and
  v-descriptors,''
\newblock {\em IEEE Transactions on Industrial Informatics}, vol. 17, no. 11,
  pp. 7456--7467, 2021.

\bibitem{qiblawey2021detection}
Yazan Qiblawey et~al.,
\newblock ``Detection and severity classification of covid-19 in ct images
  using deep learning,''
\newblock {\em Diagnostics}, vol. 11, no. 5, pp. 893, 2021.

\bibitem{lieveld2021chest}
Arthur~WE Lieveld et~al.,
\newblock ``Chest ct in covid-19 at the ed: validation of the covid-19
  reporting and data system (co-rads) and ct severity score: a prospective,
  multicenter, observational study,''
\newblock {\em Chest}, vol. 159, no. 3, pp. 1126--1135, 2021.

\bibitem{li2020ct}
Kunwei Li et~al.,
\newblock ``Ct image visual quantitative evaluation and clinical classification
  of coronavirus disease (covid-19),''
\newblock {\em European radiology}, vol. 30, pp. 4407--4416, 2020.

\bibitem{li2020clinical}
Kunhua Li et~al.,
\newblock ``The clinical and chest ct features associated with severe and
  critical covid-19 pneumonia,''
\newblock {\em Investigative radiology}, 2020.

\bibitem{hofmanninger2020automatic}
Johannes Hofmanninger et~al.,
\newblock ``Automatic lung segmentation in routine imaging is primarily a data
  diversity problem, not a methodology problem,''
\newblock {\em European Radiology Experimental}, vol. 4, no. 1, pp. 1--13,
  2020.

\bibitem{kollias2023ai}
Dimitrios Kollias, Anastasios Arsenos, and Stefanos Kollias,
\newblock ``Ai-mia: Covid-19 detection and severity analysis through medical
  imaging,''
\newblock in {\em Computer Vision--ECCV 2022 Workshops: Tel Aviv, Israel,
  October 23--27, 2022, Proceedings, Part VII}. Springer, 2023, pp. 677--690.

\bibitem{kollias2022ai}
Dimitrios Kollias et~al.,
\newblock ``Ai-mia: Covid-19 detection \& severity analysis through medical
  imaging,''
\newblock {\em arXiv preprint arXiv:2206.04732}, 2022.

\bibitem{arsenos2022large}
Anastasios Arsenos et~al.,
\newblock ``A large imaging database and novel deep neural architecture for
  covid-19 diagnosis,''
\newblock in {\em 2022 IEEE 14th Image, Video, and Multidimensional Signal
  Processing Workshop (IVMSP)}. IEEE, 2022, p. 1–5.

\bibitem{kollias2021mia}
Dimitrios Kollias et~al.,
\newblock ``Mia-cov19d: Covid-19 detection through 3-d chest ct image
  analysis,''
\newblock in {\em Proceedings of the IEEE/CVF International Conference on
  Computer Vision}, 2021, p. 537–544.

\bibitem{kollias2020deep}
Dimitrios Kollias et~al.,
\newblock ``Deep transparent prediction through latent representation
  analysis,''
\newblock {\em arXiv preprint arXiv:2009.07044}, 2020.

\bibitem{kollias2020transparent}
Dimitris Kollias et~al.,
\newblock ``Transparent adaptation in deep medical image diagnosis.,''
\newblock in {\em TAILOR}, 2020, p. 251–267.

\bibitem{kollias2018deep}
Dimitrios Kollias et~al.,
\newblock ``Deep neural architectures for prediction in healthcare,''
\newblock {\em Complex \& Intelligent Systems}, vol. 4, no. 2, pp. 119–131,
  2018.

\bibitem{frei1977image}
Werner Frei,
\newblock ``Image enhancement by histogram hyperbolization,''
\newblock {\em Computer Graphics and Image Processing}, vol. 6, no. 3, pp.
  286--294, 1977.

\bibitem{zhu2012automatic}
Yanjie Zhu et~al.,
\newblock ``Automatic segmentation of ground-glass opacities in lung ct images
  by using markov random field-based algorithms,''
\newblock {\em Journal of digital imaging}, vol. 25, no. 3, pp. 409--422, 2012.

\bibitem{maragos1987morphological}
Petros Maragos and Ronald Schafer,
\newblock ``Morphological filters--part i: Their set-theoretic analysis and
  relations to linear shift-invariant filters,''
\newblock {\em IEEE Transactions on Acoustics, Speech, and Signal Processing},
  vol. 35, no. 8, pp. 1153--1169, 1987.

\bibitem{otsu1979threshold}
Nobuyuki Otsu,
\newblock ``A threshold selection method from gray-level histograms,''
\newblock {\em IEEE transactions on systems, man, and cybernetics}, vol. 9, no.
  1, pp. 62--66, 1979.

\bibitem{yang2020chest}
Ran Yang, Xiang Li, Huan Liu, Yanling Zhen, Xianxiang Zhang, Qiuxia Xiong, Yong
  Luo, Cailiang Gao, and Wenbing Zeng,
\newblock ``Chest ct severity score: an imaging tool for assessing severe
  covid-19,''
\newblock {\em Radiology: Cardiothoracic Imaging}, vol. 2, no. 2, pp. e200047,
  2020.

\bibitem{chen2015xgboost}
Tianqi Chen, Tong He, Michael Benesty, Vadim Khotilovich, Yuan Tang, Hyunsu
  Cho, Kailong Chen, Rory Mitchell, Ignacio Cano, Tianyi Zhou, et~al.,
\newblock ``Xgboost: extreme gradient boosting,''
\newblock {\em R package version 0.4-2}, vol. 1, no. 4, pp. 1--4, 2015.

\bibitem{geurts2006extremely}
Pierre Geurts, Damien Ernst, and Louis Wehenkel,
\newblock ``Extremely randomized trees,''
\newblock {\em Machine learning}, vol. 63, pp. 3--42, 2006.

\end{thebibliography}
}

\end{document}